\documentclass[prl,floatfix,superscriptaddress,showpacs]{revtex4}
\usepackage{amsmath}
\usepackage{amsfonts}
\usepackage{amssymb}
\usepackage{graphicx}
\newcommand{\ket}[1]{\left|\, #1 \right\rangle}

\newcommand{\mean}[1]{\left\langle \,  #1 \, \right\rangle}

\begin{document}

\title{Magnetic Susceptibility as a Macroscopic Entanglement Witness}

\author{Marcin Wie\'sniak}\affiliation{Institut f\"ur Experimentalphysik, Universit\"at
Wien, Boltzmanngasse 5, A--1090 Wien, Austria}\affiliation{Instytut
Fizyki Teoretycznej i Astrofizyki Uniwersytet Gda\'nski, PL-80-952
Gda\'nsk, Poland}\affiliation{Erwin Schr\"odinger International
Institute for Mathematical Physics, Boltzmanngasse 9, A--1090 Wien,
Austria}
\author{Vlatko Vedral}\affiliation{The School of Physics and Astronomy, University of Leeds, Leeds,
LS2 9JT, United Kingdom}\affiliation{Institut f\"ur
Experimentalphysik, Universit\"at Wien, Boltzmanngasse 5, A--1090
Wien, Austria}
 \author{{\v C}aslav Brukner} \affiliation{Institut
f\"ur Experimentalphysik, Universit\"at Wien, Boltzmanngasse 5,
A--1090 Wien, Austria}\affiliation{Institut f\"ur Quantenoptik und
Quanteninformation, \"Osterreichische Akademie der Wissenschaften,
Boltzmanngasse 3, A-1090 Wien, Austria}

\date{\today}

\begin{abstract}

We show that magnetic susceptibility can reveal spin entanglement between individual
constituents of a solid, while magnetization describes their local properties. We then
show that magnetization and its variance (equivalent to magnetic susceptibility for a
wide class of systems) satisfy complementary relation in the quantum-mechanical sense.
It describes sharing of (quantum) information in the solid between spin entanglement and
local properties of its individual constituents. Magnetic susceptibility is shown to be
a macroscopic (thermodynamical) spin entanglement witness that can be applied without
complete knowledge of the specific model (Hamiltonian) of the solid.

\end{abstract}

 \maketitle

Thermodynamical properties, such as heat capacity, magnetization or
magnetic susceptibility, are normally ascribed to macroscopic
objects with the number of individual constituent of the order of
$10^{23}$. In contrast, genuine quantum features like quantum
superposition or entanglement are generally not seen beyond
molecular scales. As mass, size, complexity and/or temperature of
systems increase the observability of their quantum effects is
gradually limited by decoherence - an interaction of the system with
its environment - that turns them into classical phenomena. This
raises several questions: under which conditions can quantum
features of individual constituents of a solid have an effect on its
global properties? Can one detect existence of quantum entanglement
in a solid by observing its thermodynamical properties only? Can one
consider macroscopic properties as quantum-mechanical observables in
the sense that they obey complementary relations like position and
momentum?

The complementarity principle is the assertion that there exist
observables which are mutually exclusive in the sense that they
cannot be precisely defined simultaneously. One of them, for
example, the $z$ component of the spin-$\frac{1}{2}$ $(\sigma_z)$,
might be well defined at the expense of maximum uncertainty about
the other orthogonal directions ($\sigma_x$ and $\sigma_y$,
$\sigma$'s are respective Pauli matrices). One can speak about
sharing of (quantum) information between mutually complementary
observables~\cite{Caslav}. In the case of a qubit this can
quantitatively be described by the relation
$\langle\sigma_x\rangle^2\!+\!\langle\sigma_y\rangle^2\!+\!\langle\sigma_z\rangle^2\!\leq\!
1$, where the average is taken over an arbitrary state. When
extended to composite systems the principle of complementarity
asserts the mutual exclusiveness between entanglement and local
properties of individual constituents of the composite system. In
the case of two qubits this can be described by the relation
$\sum_{i=x,y,z}\langle\sigma_i^1\rangle^2\!+\!\langle\sigma_i^2\rangle^2\!+\!\langle\sigma^1_i\sigma^2_i\rangle^2\!\leq
3$, where the upper indices indicate qubits. The maximal value of
$3$ can be achieved, e.g., with product states (e.g.
$\langle\sigma_z^1\rangle\!=\!\langle\sigma_z^2\rangle\!=\!\langle\sigma_z^1\sigma_z^2\rangle\!=\!1$;
others are zero) for which local properties of the qubits are
well-defined, but there is no entanglement. Alternatively, their
joint properties could be well-defined at the expense of a complete
indefiniteness of the local properties (e.g. for a singlet state
$\langle\sigma_x^1\sigma_x^2\rangle\!=\!
\langle\sigma_y^1\sigma_y^2\rangle\!=\!
\langle\sigma_z^1\sigma_z^2\rangle\!=\!-1$; others are zero).

Recently, a complementarity relation was proposed~\cite{beatrix} between two macroscopic
quantities, magnetization and magnetic susceptibility along {\it one} spatial direction.
However, because entanglement between spin-$\frac{1}{2}$ systems necessarily involves
correlations at {\it different} spatial directions this cannot distinguish between
classical and quantum correlations (not a motivation of~\cite{beatrix}), which is the aim
of the present paper.

Here we will show that magnetic susceptibility, when measured along three orthogonal
spatial directions, can reveal entanglement between individual spins in a solid, while
magnetization describes their local properties. We also show that for a large class of
systems magnetization and (zero-field) magnetic susceptibility, when combined in a
particular way, satisfy a complementarity relation in the quantum-mechanical sense. This
{\it macroscopic (thermodynamical) quantum complementarity relation} describes sharing of
(quantum) information between entanglement and local properties of individual spins in
the macroscopic solid sample (in an analogy with the relation given above for two
qubits). To this end we will first prove that the sum of magnetic susceptibilities
measured along $x$, $y$ and $z$ directions is a macroscopic witness of spin entanglement
for a wide class of solid state systems. In contrast to internal
energy~\cite{wang,vb,toth,dowling,wu}, the present entanglement witness is more general
(not only valid for special materials~\cite{ghosh,cav}), can be directly measured in an
experiment, and does not require the complete knowledge of the Hamiltonian of the system.

We consider a composite system consisting of $N$ spins of an
arbitrary spin length $s$ in a lattice, which is described by a spin
Hamiltonian $H_0$. One should mention that entanglement in solids
may exist in many different degrees of freedom, such as spin,
spatial or occupation number degrees of freedom. In this letter we
will refer to the spin, assuming, moreover, that spins can be
precisely localized in sites of a lattice. Note, however, that other
degrees of freedom could also sometimes be represented in the
formalism of Pauli spin matrices~\cite{holstein}. Our results could,
therefore, also be applicable to these other scenarios.

In order to study its magnetic response properties, the solid is now put in a weak
magnetic field, say, directed along $z$-axis and of probe magnitude $B_p$ resulting in
an additional term $H_1=B_p\sum_{i=1}^{N}s_z^i$ of the Hamiltonian (here and throughout
the paper the unit $\hbar=1$ is assumed). By $s_a^i (a=x,y,z)$ we mean here an $a$-th
component of the $i$-th spin operator in the lattice. Then the Hamiltonian becomes
$H=H_0+H_1$. When the system is in its thermal equilibrium under a certain temperature
$T$, it is in a thermal state $\rho\!=\!e^{-H/kT}/Z$, where $Z\!=\!\text{Tr}(e^{-H/kT})$
is the partition function, and $k$ is the Boltzmann constant. From the partition
function one can derive all thermodynamical quantities, e.g. the magnetization
$M_z\!=\!(1/Z\beta)(\partial Z/\partial B_p)$ or the magnetic susceptibility
$\chi_z\!=\!(\partial M_z/\partial B_p)$, where $\beta\!=\!1/kT$.

It can be shown that if $[H_0,H_1]=0$, magnetic susceptibility is given by
\cite{schwabl}:
\begin{eqnarray}
\chi_z &=& \frac{1}{kT} \triangle^2(M_z) = \frac{1}{kT}(\langle
M_z^2\rangle-\langle M_z\rangle^2) \label{put} = \frac{1}{kT} \left(
\sum_{i,j=1}^N \langle s_z^i s^j_z\rangle-\left\langle \sum_{i=1}^N
s_z^i \right\rangle^2 \right),
\end{eqnarray}
where $\triangle^2(M_z)$ is variance of the magnetization.

Microscopically, magnetic susceptibility is, in fact, a sum over all microscopic spin
correlation functions $\langle s_z^i s^j_z \rangle$ for the sites $i$ and $j$. The above
is a very important relation as it connects a macroscopic quantity to its microscopic
roots in the form of the two-site correlation functions. Note, however, that the nonzero
value of the correlation function does not necessarily imply the existence of
entanglement. What we need, loosely speaking, are sufficiently strong correlations in
all three orthogonal spatial directions and they need to be combined in a specific way
to reveal spin entanglement. This is the reason why we will now study the sum of
magnetic susceptibilities $\chi_x$, $\chi_y$ and $\chi_z$ for weak probe fields aligned
along three orthogonal directions.

We now show that the expression $\chi_x+\chi_y+\chi_z$ is an
entanglement witness. Entanglement witnesses in general are
observables which (by our convention) have positive expectation
values for separable states and negative ones for some specific
entangled states~\cite{horodecki}. The proof is based on the method
of entanglement detection using the uncertainty
relations~\cite{hofmann}. For any separable state of $N$ spins of
length $l$, (for any classical mixture of the products states, each
appearing with probability $w_n$: $\rho=\sum_n w_n \rho^1_n \otimes
\rho^2_n \otimes ... \otimes \rho^N_n$), one has
\begin{equation}
\label{total} \bar{\chi} \equiv \chi_x+\chi_y+\chi_z \geq \frac{N
s}{kT}.
\end{equation}

We will now prove this inequality for product states of the spins and then the general
result will follow directly due to convexity of separable states. Note that for an
arbitrarily state of spin-$s$ particle one has $\mean{(s_x)^2}+$$\mean{(s_y)^2}+ $$
\mean{(s_z)^2}$$ = $$ s(s+1)$ and $\langle s_x \rangle^2 + $$ \langle s_y \rangle^2 + $$
\langle s_z \rangle^2 $$ \leq s^2 $. If the thermal state was
 actually a product one of $N$ spins, the variance of magnetization
would be the sum of variances of individual spins:
$\bar{\chi}=(1/kT)$ $ \left[\triangle^2(M_x) + \triangle^2(M_y) +
\triangle^2(M_z) \right]$ $ =(1/kT) \sum_i \left[\triangle^2 (s^i_x)
+ \triangle^2 (s^i_y) + \triangle^2 (s^i_z) \right] \geq
(1/kT)(s(s+1)-s^2) = Ns/(kT)$. Note that this bound is also valid in
the general case of separable states due to the convexity of the
mixture:
\begin{eqnarray}
\bar{\chi}&=&(1/kT)(\triangle^2(M_x)_\rho+\triangle^2(M_y)_\rho+
\triangle^2(M_z)_\rho)\nonumber\\
 &\geq &(1/kT)\sum_n w_n\sum_i \left[\triangle^2 (s^i_x)_n + \triangle^2 (s^i_y)_n +
\triangle^2 (s^i_z)_n \right]\nonumber\\ &\geq & Ns/(kT),
\end{eqnarray}
where index $n$ denotes the $n$-th subensemble in the mixture, and
$\triangle^2(X)_\rho$ is a variance of an observable $X$ taken in a
state $\rho$. The inequality (\ref{total}) is saturated for any pure
state which has a maximal magnetic quantum number $m$ with respect
to any direction (i.e. $\ket{j,m=j})$. The other extreme case of
$\bar{\chi}=0$ can be achieved for a singlet state of $N$ spins
where all three variances are equal to zero.

\begin{figure}[h]
\centering
\includegraphics[width=6cm]{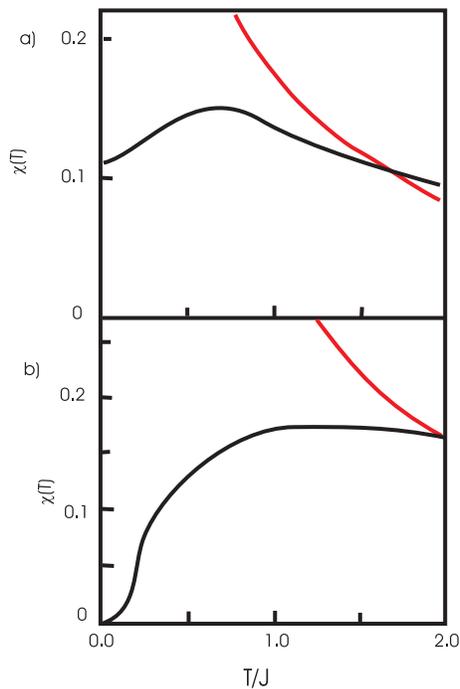}
\caption{Detection of entanglement in the xxx Heisenberg
spin-$\frac{1}{2}$ (a) and spin-1 chains (b). The black solid curves
are the theoretical curves from Ref.~\cite{xiang} and represent the
temperature dependence of the zero-field magnetic susceptibility
$\chi(T)$ per particle in the spin-$\frac{1}{2}$ (a) and spin-1
chains (b). The red solid curves are from our work and represent the
right-hand side of inequality~(\ref{iso3d}). Mathematically, they
represent hyperbolae $1/(6T)$ (a) and $1/(3T)$ (b). The critical
temperatures below which entanglement exists in the chains are $T_c
=1.6J$ for spins-$\frac{1}{2}$ and $T_c=2J$ for spins-1 (The data in
\cite{xiang} were given only for $T\leq 2J$).} \label{xiang}
\end{figure}

Therefore, if $\chi_x+\chi_y+\chi_z  < \frac{Ns}{kT}$, the solid state system contains
entanglement between individual spins. It is important to note that all susceptibilities
should be taken for zero-fields $B_p$ to ensure that they are measured for the same
thermal state (for the same reason no quantum phase transition at the points of the
measurements is assumed). Because the measurement of the magnetic susceptibility has
been an experimental routine for long time, it is clear that the present approach is an
experimentally efficient method for detecting macroscopic spin entanglement. It might be
of particular importance when there is only partial knowledge of systems's Hamiltonian
and one thus has to rely more on experiment. In what follows we will demonstrate the
efficiency of the method using an exactly solvable model.

Suppose that the symmetry of the system is such that magnetic susceptibility is equal in
all three directions $\chi_x\!=\!\chi_y\!=\!\chi_z$. This is the case, for example, for
the Heisenberg spin lattices with isotropic, but in general inhomogeneuous coupling
constant $J_{ij}$: $H_0=\sum_{i,j}J_{ij}\vec{s}^i\vec{s}^j$ (here the summation does not
need to be constrained to nearest-neighbour interactions only). The entanglement
criterion now reads as follows:
\begin{equation}
\label{iso3d} \chi_z < \frac{1}{3} \frac{N s}{k T}.
\end{equation}
We apply it to investigate existence of entanglement in the infinite
xxx Heisenberg chains of spins-1/2 (Fig. 1a) and spins-1 (Fig. 1b),
described by a Hamiltonian $H_0=\sum_{i}\vec{s}_i\vec{s}_{i+1}$, at
various temperatures. We use the results of Ref.~\cite{xiang} where
the thermodynamic properties of the Heisenberg spin chains are
obtained by the transfer-matrix renormalization-group method. The
right-hand side of~(\ref{iso3d}) is represented by the red solid
lines in Fig.~\ref{xiang}. The values of magnetic susceptibility
left to the intersections point of the red and the theoretical
curves cannot be explained without entanglement. The critical
temperatures are $T_c =1.6J$ for spins-$\frac{1}{2}$ and $T_c=2J$
for spins-1.

It is commonly believed that one reconstructs the physics of classical spins in the
limit of infinitely large spins as suggested by the limit of the commutation relation
for normalized spins $S_i \equiv \frac{s_i}{s}$ ($i=x,y,z$): $\lim_{s\rightarrow \infty}
[S_x,S_y]=\lim_{s\rightarrow \infty}\frac{1}{s} S_z\rightarrow 0$. Our result shows,
however, that the larger the spins are, the higher the critical temperature is below
which entanglement is present.  Perhaps one possible explanation of this effect is that
the longer spins produce entanglement of higher dimension than the one produced by
shorter spins. Thus, for the same coupling strength $J$ and temperature the longer spins
can develop larger amount of entanglement than the shorter spins, which might then
persist at higher temperatures. The dependence of an amount of entanglement on
dimensionality of subsystems was studied in Ref.~\cite{dennison}.

It is interesting to compare the threshold temperature for the
spin-$\frac{1}{2}$ chain with the estimates based on other
macroscopic entanglement witness such as internal energy. In
Ref.~\cite{wz,vb} it was shown that the internal energy can reveal
the entire bipartite entanglement between neighboring spins as
measured by the concurrence~\cite{wootters}. Using these and the
results from~\cite{xiang} we obtain that the concurrence vanishes
below the threshold temperature of $0.795J$. The higher value of
$1.6J$ as revealed by magnetic susceptibility can be explained by
the fact that it, in contrast to internal energy, may also detect
bipartitive entanglement between non-neighboring sites and
multipartitive entanglement.

We now turn to the derivation of a macroscopic quantum complementarity relation. We
first note that the sum of the squares of magnetizations along three orthogonal
directions satisfies the relation: $\langle \vec{M}\rangle^2\equiv \langle
M_x\rangle^2+\langle M_y\rangle^2+\langle M_z\rangle^2 \leq N^2s^2$. This describes the
complementarity between properties of individual spins in a solid, in analogy with the
$\langle\sigma_x\rangle^2\!+\!\langle\sigma_y\rangle^2\!+\!\langle\sigma_z\rangle^2\!\leq\!
1$ for a single qubit. If one of the observables in the sum, for example $\langle M_z
\rangle^2$, takes its maximal value of $N^2s^2$ (e.g. in state
$|j\!=\!Ns,m\!=\!Ns\rangle$ where $j$ is angular and $m$ magnetic quantum number), the
other two have to vanish. For the purposes of further discussion we need the following
relation between $\langle \vec{M}\rangle^2$ and $\langle \vec{M}^2\rangle \equiv \langle
\vec{M_x}^2\rangle + \langle \vec{M_y}^2\rangle + \langle \vec{M_z}^2\rangle$:
\begin{equation}
\langle \vec{M}^2\rangle\geq\left(\frac{Ns+1}{Ns}\right)\langle
\vec{M}\rangle^2. \label{srce}
\end{equation}
Here follows the proof. Let us denote by $\ket{j,m}$ the joint
eigenstates of $\vec{M}^2$ with eigenvalues $j(j+1)$ and $M_z$ with
eigenvalues $m$. Note that both sides of (\ref{srce}) are invariant
under rotations in the three-dimensional space. Thus, for any given
state we can choose such a coordinate system that $\langle M_x
\rangle \! =\!\langle M_y \rangle\!=\!0$ and consequently
$\langle\vec{M}\rangle^2\!=\!\langle M_z \rangle^2$. Let us now
define a new operator $K$ such that $K\ket{j,m}\!=\!j\ket{j,m}$.
Given that $p_{jm}$ are probabilities of finding the system in any
of the states $\ket{j,m}$, we have $\langle M_z
\rangle\!=\!\sum_{j,m}p_{jm} m \! \leq \! \sum_{j,m} p_{jm}j \!=\!
\langle K \rangle$. To complete the proof we use
$\langle\vec{M}^2\rangle\!=\!\langle K (K+1) \rangle$ and the fact
that $Ns\! \geq \!j$ implies $Ns \langle K \rangle \!\geq \!\langle
K^2 \rangle$. Thus we have
\begin{equation}
\langle\vec{M}^2\rangle-\left(\frac{Ns+1}{Ns}\right)\langle\vec{M}\rangle^2
\geq\langle K (K+1) \rangle-\left(\frac{Ns+1}{Ns}\right)\langle
K\rangle^2\geq
\left(\frac{Ns+1}{Ns}\right)\triangle^2\left(K\right)\geq 0.
\end{equation}

We now exploit Eq.~(\ref{put}) and (\ref{srce}) to derive a
macroscopic quantum complementarity relation:
\begin{equation}
\label{comp} \underbrace{1-\frac{kT\bar{\chi}}{Ns}}_{non-local \mbox{ } properties} +
\underbrace{\frac{\langle \vec{M}\rangle^2}{N^2s^2}}_{local \mbox{ } properties} \leq 1.
\end{equation}
The left-hand side of inequality~(\ref{comp}) can be divided into
two parts: $Q \!\equiv\! 1-\frac{kT\bar{\chi}}{Ns}$ and
$P\equiv\frac{\langle\vec{M}\rangle^2}{N^2s^2}$. While $P$ describes
the local properties of individual spins, $Q$ is associated with
quantum correlations between spins in a solid. This is because $Q$
is proportional to two-site spin correlations for three orthogonal
directions (three mutually non-commuting observables) and its
positive value implies the existence of entanglement (see
Eq.~(\ref{total})). In the extreme case of a product state of $N$
spins all aligned along the same direction (e.g.
$|j\!=\!Ns,m\!=\!Ns\rangle$), their local properties are well
defined ($P\!=\!1$) at the expense of no entanglement ($Q\!=\!0$).
In the other extreme case the state of the systems is highly
entangled. Then non-local properties are maximally exhibited
($Q=1$), at the expense of $P\!=\!0$. In general, the relation
(\ref{comp}) describes partial quantum information sharing between
local and non-local properties of spins.

\begin{figure}
\includegraphics[width=7cm]{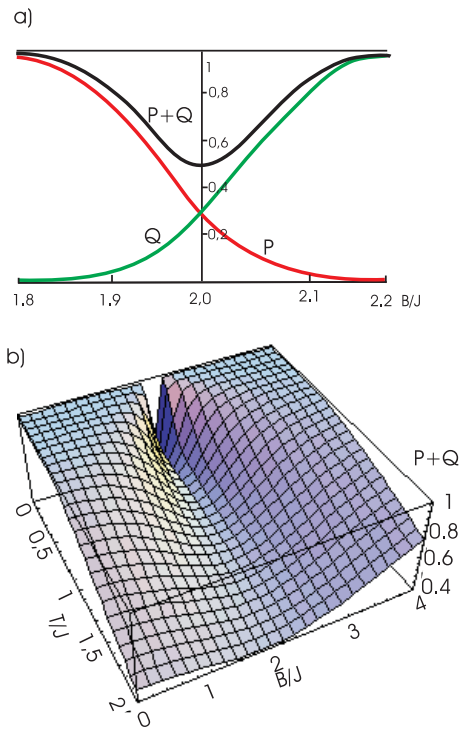}
\caption{Macroscopic quantum complementarity relation between local ($P$) and non-local
$(Q)$ properties. (a) The plot of $Q=1-2kT\bar{\chi}/N$ (red),
$P=4\langle\vec{M}\rangle^2/N^2$ (green) and its sum $P+Q$ (black) for a chain of
antiferromagnetically coupled spin-$\frac{1}{2}$ pairs (dimers) versus the magnetic
field $B/J$. The temperature is taken to be $T=0.1J$. (b) The plot of $P+Q$ as a
function of magnetic field $B/J$ and temperature $T/J$. Under all temperatures and
values of magnetic field the complementarity relation $P+Q \leq 1$ is satisfied (see
text for further discussion).} \label{graph}
\end{figure}

To illustrate the complementarity relation (\ref{comp}) we analyze
a chain of antiferromagnetically coupled spin pairs - dimers -
which are themselves uncoupled. This is a correct model for,
e.g., Copper Nitrates and many organic radicals. The Hamiltonian
in an external magnetic field of magnitude $B$ is given by
\begin{equation}
H_0=J \sum_j \vec{\sigma}^{2j} \cdot \vec{\sigma}^{2j+1} + B \sum_j \sigma^j_z.
\end{equation}
The plot of $P$, $Q$ and their sum $P+Q$ as a function of the magnetic field $B$ and
temperature $T$ is given in Fig.~(\ref{graph}). For $B=0$, the singlet is the ground
state and the triplets are the degenerate excited states. For a higher value of $B$,
however, the triplet states split and the gap between the singlet and first excited
state $|--\rangle$ ($\sigma_z\ket{-}=-\ket{-}$) decreases. Therefore, in a thermal state
at a given temperature as $B$ is increased, the definiteness of non-local properties $Q$
decreases because increasingly larger triplet component will be mixed with the singlet
\cite{arnesen,Nielsen}. On the other hand, as $B$ increases the spins tends to orient
themselves all parallel to the field, which results in higher values of magnetization
and thus $P$, in agreement with the complementarity relation (see Fig.~(\ref{graph})).
Increasing $T$ generally decreases $P+Q$ as thermal mixing has a destructive character
both to $Q$ and $P$. Note, however, that at all temperatures and all values of magnetic
field the relation $P+Q \leq 1$ is satisfied.

In conclusions, we show that magnetic susceptibility is an entanglement witness for a
wide class of systems. While magntetization describes local properties of individual
constituents of a solid, the magnetic susceptibility specifies its spin entanglement. We
show that magnetization and magnetic susceptibility satisfy a quantum complementarity
relation. One of the quantities can thus increase only at the expense of a decrease in
the other. This shows quantum information sharing in macroscopic quantum systems, such
as solids. In future, it will be interesting to investigate this feature at critical
points where (quantum) phase transitions occur. It can be seen from our plot at $T=0$
(Fig.~\ref{graph}a) that a sudden increase in non-local properties ($Q$) at the quantum
phase transition point $B/J=2$ must be accompanying with a corresponding sudden decrease
in $P$. Otherwise, the complementarity relation would be violated.

Our results are not only relevant for fundamental research but also for quantum
information science as they give the critical values of physical parameters (e.g. the
high-temperature limit) above which one cannot harness quantum entanglement in condensed
matter systems as a resource for quantum information processing.

{\v C}.B. and M.W. were supported by the Austrian Science Foundation (FWF) Project SFB
1506. M.W. is supported by the Erwin Schr\"odinger Institute in Vienna and the
Foundation for Polish Science (FNP). {\v C}.B. thanks the European Commission (RAMBOQ).
V. V. thanks European Union and the Engineering and Physical Sciences Research Council
for financial support. {\v C}.B. and V.V. thank the British Council in Austria. The
authors would like to thank B. Hiesmayr, A. Fereirra, and J. Kofler on very useful
discussions and comments. The work was a part of Austrian-Polish Collaboration Programme
``Quantum Information and Quantum Communication V".

\end{document}